# Dielectric Relaxation and Phase Transition at Cryogenic Temperatures in $0.65[Pb(Ni_{1/3}Nb_{2/3})O_3]$-$0.35PbTiO_3$ Ceramics

Satendra Pal Singh, Akhilesh Kumar Singh and Dhananjai Pandey

*School of Materials Science & Technology, Institute of Technology,*

*Banaras Hindu University, Varanasi-221 005, India*

S. M. Yusuf

*Solid State Physics Division, Bhabha Atomic Research Centre,*

*Mumbai-400 085, India*

## Abstract

Dielectric measurements on $0.65[Pb(Ni_{1/3}Nb_{2/3})O_3]$-$0.35PbTiO_3$ ceramic in the temperature range 90K to 470K shows a relaxor ferroelectric transition around 350K with a Vogel-Fulcher freezing temperature of 338K and appearance of a non-ergodic relaxor ferroelectric phase of tetragonal structure at room temperature. This non-ergodic phase reenters into the relaxor state at low temperatures as evidenced by the appearance of a frequency dependent anomaly in the imaginary part of the dielectric constant around 160K, similar to those reported in other relaxor ferroelectric based morphotropic phase boundary ceramics. The polarization relaxation time for the 160K anomaly also follows Vogel-Fulcher type temperature dependence. Temperature dependent magnetization measurements show that this low temperature anomaly is not linked with any magnetic transition. Elastic modulus and low temperature x-ray diffraction (XRD) measurements reveal a tetragonal to monoclinic phase transition around 225K. It is argued that the low temperature dielectric dispersion around 160K results from the freezing of mesoscopic conformally miniaturized monoclinic domains formed inside the parent tetragonal domains below the structural phase transition temperature of 225K.

## 1. Introduction:

Currently there is enormous interest in multiferroic materials, especially the magnetoelectrics, because of their potential applications in advanced sensor and actuator technology [1]. $Pb(Ni_{1/3}Nb_{2/3})O_3$ (PNN) is a multiferroic showing a relaxor ferroelectric transition around 153K [2] and an antiferromagnetic transition at very low temperatures (~5K) [3]. The phase diagram of the solid solution of PNN with $PbTiO_3$, i.e. $(1-x)[Pb(Ni_{1/3}Nb_{2/3})O_3]$ -$xPbTiO_3$ (PNN-xPT), exhibits a morphotropic phase boundary (MPB) similar to the well known piezoelectric ceramics, like $(1-x)[Pb(Mg_{1/3}Nb_{2/3})O_3]$-$xPbTiO_3$ (PMN-xPT) [4],$(1-x)[Pb(Zn_{1/3}Nb_{2/3})O_3]$-$xPbTiO_3$ (PZN-xPT) [5] and $Pb(Zr_xT_{1-x})O_3$ (PZT) [6]. Recently, low temperature studies on single crystals of PZN-x PT [7, 8] and PMN-xPT [9, 10] have revealed an unexpected frequency dependent dielectric anomaly at cryogenic temperatures well below the main relaxor ferroelectric transition temperature. As yet, there is no satisfactory explanation for the physical origin of this dielectric anomaly, although two different models have been proposed. One of the models is based on the thermal agitation of local polarization fluctuations induced by chemical heterogeneities [7, 11] while the other assumes the presence of fractal clusters inside the normal ferroelectric domains as the main cause of the dielectric anomaly at low temperatures [9, 12]. It has been proposed by Bao *et al* [8] and Lente *et al* [10] that this low temperature dielectric anomaly is not linked with a structural phase transition. However, no structural studies have been reported by these workers in corroboration of their proposition. We present here the results of a combined dielectric, magnetic, elastic modulus, x-ray diffraction (XRD) and polarization study on a multiferroic MPB ceramic, PNN-xPT with x = 0.35, to demonstrate not only the universality of the phenomenon of the low temperature dielectric relaxation in all such relaxor based MPB ceramics including

PNN-xPT, but also point towards an indirect role of the low temperature structural phase transition in this phenomenon, overlooked so far in the previous studies.

**2. Experimental:**

Analytic reagent (AR) grade chemicals $NiCO_3.2Ni(OH)_2.4H_2O$, $Nb_2O_5$, $PbCO_3$, and $TiO_2$ with minimum assay of 99% or more were used to synthesize PNN-xPT. Stoichiometric mixtures of various ingredients were ball milled (Restch GmbH & Rheinische, Germany) using zirconia jars and zirconia balls with AR grade acetone as the mixing media. A columbite precursor $NiNb_2O_6$ was first prepared by calcining the mixture of stoichiometric amounts of $NiCO_3.2Ni(OH)_2.4H_2O$ and $Nb_2O_5$ at 1050°C for 6 hours. Stoichiometric amount of $TiO_2$ was then mixed with $NiNb_2O_6$ and the mixture was calcined at 1050°C for another 6 hours to obtain $[(1-x)/3]NiNb_2O_6$-$xTiO_2$ (NNT) precursor. This NNT precursor was finally mixed with stoichiometric amount of $PbCO_3$ and calcined at 800°C for 6 hours. Cold compaction of the calcined powder was done using steel die of 12-mm diameter and an uniaxial hydraulic press at an optimized load of 65 kN. 2% polyvinyl alcohol (PVA) solution in water was used as a binder. The green pellets were kept at 500°C for 10 hours to burn off the binder material and then sintered at 1150°C for 6 hours in a sealed alumina crucible with controlled PbO atmosphere. The density of the sintered pellets was higher than 98% of the theoretical value. Fired-on silver paste was used for electroding the sintered pellets. The dielectric measurements were carried out using a Novocontrol Alpha-A High Performance Frequency Analyser. The sample temperature was varied using programmable temperature controller at a cooling and heating rate of 1K/min. The dielectric measurements above (300-470K) and below (300-90K) room temperature were carried out in two different set-ups. For piezoelectric resonance and antiresonance frequency measurements, the electroded pellets were poled at a dc field of 35kV/cm at 323K. Piezoelectric resonance and antiresonance frequencies ($f_r$ and $f_a$) were measured using a Schlumberger Impedance/Gain-Phase Analyzer SI 1260. For x-ray characterization, the sintered pellets were crushed into fine powders and then annealed at 500°C for 10 hours to remove the strains introduced during crushing. XRD measurements were carried out using an 18kW rotating anode (Cu) based Rigaku powder diffractometer (operating in the Bragg-Brentano geometry) fitted with an ultra low temperature attachment and graphite monochromator in the diffracted beam. Hysteresis loop measurements were performed using a modified Sawyer-Tower circuit and an Agilent 54624A storage oscilloscope. The dc magnetization measurements were carried out using a 12 Tesla commercial (Oxford Instruments) vibrating sample magnetometer (VSM) as a function of temperature and magnetic field. Temperature dependence of magnetization was carried out in the warming cycle over 5-315K under 100 Oe external field after first cooling the sample from 315K to 5K in the same field.

**3. Results and Discussion**

**3.1 Relaxor Ferroelectric and Magnetic Transitions:**

The composition PNN-0.35PT has a tetragonal structure at room temperature and is close to the MPB which occurs at x ≈ 0.36 [13, 14]. Fig.1 depicts the variation of the real [$\varepsilon^{/}(T)$] and imaginary [$\varepsilon^{//}(T)$] parts of the dielectric constant as a function of temperature at various frequencies in the range 100Hz to 10 kHz for this composition. It can be seen from this figure that a diffuse dielectric peak in the real part of the dielectric constant [$\varepsilon^{/}(T)$] appears around 350K. The imaginary part of the dielectric constant [$\varepsilon^{//}(T)$], on the other hand, shows two peaks, first around 342K and another around 160K [see Fig 1 (a) and Fig. 1(b)]. No anomaly in $\varepsilon^{/}(T)$ corresponding to the low temperature peak in $\varepsilon^{//}(T)$ is apparent in Fig. 1(b) .The temperatures $T^{/}_{m1}$ and $T^{//}_{m1}$, corresponding to the high temperature peaks in $\varepsilon^{/}(T)$ and $\varepsilon^{//}(T)$, respectively, shift to higher temperature side on increasing the measuring frequency. This is shown more clearly in the inset (i) of Fig. 1(a) for $\varepsilon^{/}(T)$. It is also evident that the peak temperature $T^{//}_{\rm m1}$ does not coincide with $T^{/}_{\rm m1}$; instead $T^{//}_{\rm m1} < T^{/}_{\rm m1}$. All these features suggest relaxor ferroelectric nature of the dielectric peak occurring around 350K [15, 16]. The second broad peak in $\varepsilon^{//}(T)$ below $T^{/}_{\rm m1}$ and $T^{//}_{\rm m1}$, in the temperature range 160K to 170K, also shows frequency dependent shifts. This frequency dependent shift of $T^{//}_{\rm m2}$ at cryogenic temperatures is similar to that reported in PZN-xPT [7, 8] and PMN-xPT [9, 10] systems. The main Curie peak shows a thermal hysteresis of ~ 8K at 10kHz during heating and cooling cycles of measurements [see inset (ii) of Fig. 1(a)]. The $T^{//}_{\rm m2}$, on the other hand, remains nearly the same during heating and cooling cycles [see inset to Fig. 1(b)].

The temperature dependence of relaxation time ($\tau$), as obtained from the $\varepsilon^{//}(T)$ data, was modeled using Arrhenius and Vogel-Fulcher type relations:

$$\tau = \tau_o \exp(E_a/KT) \qquad \text{(Arrhenius)} \qquad (1)$$

$$\tau = \tau_o \exp[E_a/K(T-T_{VF})] \qquad \text{(Vogel-Fulcher)} \qquad (2)$$

Here, $\tau$ is the relaxation time, $\tau_o$ is the pre-exponential factor, $E_a$ is the activation energy and $T_{VF}$ is the Vogel- Fulcher freezing temperature. Arrhenius type behaviour can be ruled out for PNN-0.35PT since the fit between $\ln\tau$ vs $1/T$ is not linear for both the high temperature and the low temperature dielectric anomalies, as can be seen from the insets to Fig. 2(a) and (b). The Vogel-Fulcher law, on the other hand, gives very satisfactory fit as shown in Fig. 2(a) and 2(b) for the high and low temperature peaks in $\varepsilon^{//}(T)$, respectively. The least square fitted parameters are: $E_a = 2.292\times 10^{-3}$ eV and $3.15\times10^{-2}$ eV, $T_{VF} \approx$ 338K and 136K and $\tau_o = 3.078\times10^{-7}$ sec and $4.547\times10^{-10}$ sec, respectively, for the high and low temperature peaks.

The $T_{VF}$ is usually regarded as the temperature at which an ergodic relaxor ferroelectric phase transforms into a non-ergodic relaxor state due to the divergence of the longest relaxation time associated with the polarization fluctuations [17]. In the ergodic relaxor ferroelectric state of a related MPB system, PMN-xPT, the powder diffraction profiles exhibit a pseudocubic structure whereas in the non-ergodic relaxor state, splitting of x-ray powder diffraction lines have been reported [18, 19] for $x \geq 0.10$. In the pure unalloyed PMN, one usually does not observe [20] such splittings even in the non-ergodic phase. Since the $T_{VF1}$ = 338 K of PNN-0.35PT is higher than the room temperature, a non ergodic relaxor phase is expected at room temperature. This non-ergodic relaxor phase exhibits characteristic tetragonal splittings of cubic perovskite peaks in the room temperature powder diffraction data (see Fig. 5) similar to PMN-xPT for $x \geq 0.10$. The observation of the second freezing temperature $T_{VF2}$ = 136K seems to suggest that the room temperature nonergodic relaxor phase re-enters into an ergodic relaxor phase on lowering the temperature below $T_{VF1}$ giving rise to yet another non-ergodic relaxor phase at $T < T_{VF2}$.

In multiferroics like $Pb(Fe_{1/2}Nb_{1/2})O_3$ (PFN) [21] and $YMnO_3$ [22], anomalies in the dielectric constant have been reported at the antiferromagnetic transition temperature due to a coupling between the ferroelectric and antiferromagnetic order parameters. In order to explore if the low temperature anomaly in the dielectric constant around 160K in PNN-0.35PT is also linked with some magnetic transition, we carried out magnetization measurements as a function of temperature at a magnetic field of 100 Oe. The variation of magnetic susceptibility ($\chi$), obtained from the measured magnetization, with temperature is shown in Fig. 3. There is no evidence for any anomaly around 160K in the magnetic susceptibility vs. temperature plot. The $\chi$ vs. temperature plot shows a sharp rise at low temperatures ( < 50K) similar to that reported in pure PNN [3].Magnetization measurements for PNN is also included in Fig. 3 for comparison. It may therefore be concluded that the anomaly in the imaginary part of the dielectric constant is not linked with any magnetic transition. $1/\chi$ vs. temperature plots for PNN-0.35PT and PNN depicted in the inset to Fig.3 show negative Curie-Weiss temperatures ($T_o$), indicative of antiferromagnetic correlations. The antiferromagnetic transition is known to occur below 5K in pure PNN [3] and the situation seems to be similar even in PNN-0.35PT.

**3.2 Evidence for a Low Temperature Structural Phase Transition:**

Recently, tetragonal compositions of PZT [23, 24], PMN-xPT [25] and PZN-xPT [26] close to the MPB have been reported to transform to monoclinic phases below room temperature. Ragini *et al* [23] and Singh *et al* [19] have shown that these phase transitions between the tetragonal and monoclinic phases in PZT and PMN-xPT are accompanied with a imperceptibly weak anomaly in the dielectric constant of unpoled samples but a very pronounced anomaly in the elastic modulus. In order to explore the existence of such a tetragonal to monoclinic phase transition in PNN-0.35PT below room temperature, we therefore carried out piezoelectric resonance and antiresonance frequency measurements in the planar coupling mode [6] as a function of temperature, which provide the elastic compliance ($S_{11}^E$).The $S_{11}^E$ can be calculated from the measured resonance frequency ($f_r$) and other material constants using the following relationship [6].

$$\frac{1}{S_{11}^E} = \frac{\pi^2 d^2 f_r^2 (1-\sigma^E)\rho}{\eta^2}, \qquad (3)$$

where $S_{11}^{E}$ is the elastic compliance at constant field, d is the diameter of the pellet, $\sigma^{E}$ (= 0.31) is the Poisson's ratio, $\eta$ = 2.05 and $\rho$ is the pellet density. Fig.4 depicts the variation of the inverse of the elastic compliance i.e. elastic modulus (E), with temperature. The elastic modulus of normal solids, which expand on heating, should increase with decreasing temperature. However, in our case it first decreases with decreasing temperature, which is an anomalous behaviour, and may happen due to some lattice instability near a structural phase transition, as has been shown by Ragini *et al* [23] in PZT and Singh *et al* in PMN-xPT [19] ceramics. The elastic modulus after decreasing upto 225K, starts rising on lowering the temperature below 225K, as expected for a normal solid. This suggests that a low temperature structural phase transition occurs around 225K in PNN-0.35PT similar to the low temperature transitions reported recently in PZT [23, 24], PMN-xPT [25] and PZN-xPT [26]. While the real part of the dielectric constant $\varepsilon^{/}(T)$ below room temperature varies continuously with decreasing temperature without showing any clear anomaly at ~ 225K, when we plotted $1/\varepsilon^{/}(T)[d(\varepsilon^{/}(T))/dT]$ with temperature, it shows a dip around 225K as shown in the inset to Fig.4. This further corroborates a phase transition occurring around 225K.

Low temperature XRD studies confirm the existence of a structural phase transition below 225K. Fig.5 depicts the evolution of the 200, 220 and 222 pseudocubic reflections with temperature. The 200 and 220 pseudocubic reflections form a doublet, while 222 is a singlet for the tetragonal structure, whereas for the rhombohedral structure, 200 is a singlet, and the 220 and 222 reflections are doublets. The doublet nature of 200 and 220 peaks and the singlet character of the 222 peak in the XRD data at 300K confirms the tetragonal structure at room temperature. On lowering the temperature, the width of the 222 peak starts increasing below 225K, as can be inferred from a comparison of the width of the 222 profiles at $T \geq 240$K with those at $T \leq 210$K (see Fig.5). This clearly shows that the 222 peak is no longer a singlet at $T \leq 210$K, which in turn suggests the appearance of a non-tetragonal phase in the temperature range 210K $\leq T <$ 240K. The low temperature phase cannot be rhombohedral, since the 200 peak does not become a singlet at $T \leq 210$K. Thus both the elastic modulus and XRD studies reveal a low temperature structural phase transition occurring around 225K in the tetragonal PNN-0.35PT composition.

In order to determine the structure of the low temperature phase, we carried out Rietveld analysis of the powder x-ray diffraction data of PNN-0.35PT recorded at various temperatures in the range 300 to 15K using Fullprof package [27]. The Rietveld analysis confirms the tetragonal structure at room temperature as can be seen from the satisfactory fits between the observed and calculated profiles of 200, 220 and 222 pseudocubic reflections, obtained after full pattern refinement, shown in Fig. 6(a). A similar refinement, using the 15K data, gives very poor fit showing the inadequacy of the tetragonal structure (P4mm space group) in explaining the observed diffraction data [see Fig. 6(b)]. The eighth order Landau theory for ferroelectric transitions in perovskites predicts three monoclinic phases of $M_A$, $M_B$, $M_C$ types with Cm, Cm and Pm space groups [28]. Vanderbilt and Cohen [28] distinguish between $M_A$ and $M_B$ types for the Cm space group on the basis of the values of polarization components, **Px**, **Py** and **Pz**, along the pseudocubic axes. Keeping in view the recent discovery of the monoclinic phases in PZT [29, 30], PMN-xPT [31] and PFN-xPT [32], we have considered the two monoclinic phases with Cm and Pm space groups in our refinements. However, the fit was not very satisfactory. As a next step, we were led to consider the coexistence of tetragonal and monoclinic phases in our refinements. While the fit between the observed and calculated profiles becomes quite satisfactory after considering the coexistence of the two phases [see Fig. 6(c) and (d)], it was rather difficult to distinguish between the Cm + tetragonal and the Pm + tetragonal models purely on the basis of the visual observation of the fits or the various agreement factors. We therefore adopted the DW statistics [33] and Prince's criterion [32, 34] to make a choice between the two models. It was found that both the criterion favour the coexistence of the monoclinic phase in the Pm space group with the tetragonal phase. The overall fit between the observed and calculated profiles for the coexistence of the monoclinic phase in the Pm space group and the tetragonal phase in the P4mm space group is shown in Fig.7 for the 2θ range 20- 120 degrees. Table 1 lists the refined structural parameters. The weight fractions of the monoclinic and tetragonal phases at 15K are 0.60 and 0.40. With increasing temperature, the tetragonal phase fraction increases but a small portion of the monoclinic phase survives upto room temperature. This wide coexistence of the two phases is a signature of the first order character of the phase transition occurring at 225K with strong metastability effects.

### 3.3 Origin of the Low Temperature $\varepsilon^{//}$ (*T*) peak:

Since the structural phase transition temperature (225K) is significantly higher than the temperature (160K) at which the second anomaly in $\varepsilon^{//}(T)$ is observed, one may be tempted to rule out any linkage of the low temperature tetragonal to monoclinic phase transition with the appearance of the dielectric relaxation peak around 160K. The question now arises as to why the ergodic relaxor state stable above $T_{VF1}$, after having transformed into a non-ergodic tetragonal phase below $T_{VF1} \approx$ 338K, reenters into an ergodic relaxor ferroelectric phase after the tetragonal to monoclinic phase transition. One possible explanation for this could be the formation of miniaturized monoclinic domains within each tetragonal domain. Evidently, the monoclinic domains will be much smaller than the individual tetragonal domains. We believe that these domains are of mesoscopic sizes and show dielectric relaxation below the tetragonal to monoclinic phase transition temperature due to thermal fluctuations, as in superparaelectrics. This relaxation of the mesoscopic domains give rise to the low temperature dielectric relaxation peak. This explanation in terms of the dynamics of the mesoscopic monoclinic domains can also explain as to why the diffraction profiles do not exhibit splittings characteristic of the monoclinic phase, both in the laboratory and synchrotron XRD data [35]. The Scherrer broadening due to small mesoscopic size monoclinic domains apparently masks the characteristic peak splittings.

Evidence for domain fragmentation model is provided by the temperature dependent polarization hysteresis loop measurements. Hysteresis loops measured at a constant field of 12kV/cm at various temperatures is shown in Fig.8. At this constant electric field, saturated hysteresis loop was observed upto 225K only. Below this temperature, the hysteresis loop is not saturated and the loop start shrinking and finally become extremely slim at low temperatures (123K). However, on increasing the field to higher values, we could get saturated hysteresis loops below 225K also and even below 123K as shown in Fig. 9 for $T$ = 123K at 22kV/cm. The variation of remanent polarization ($P_r$) and coercive field ($E_c$) with temperature, obtained from the hysteresis loop data at constant electric field and from the saturated hysteresis loop data obtained with higher fields, is shown in Fig.10. It is evident from this figure that the remanent polarization ($P_r$) obtained at constant field shows a maximum around 225K and decreases for $T$ < 225K. Remanent polarization ($P_{rS}$) obtained from saturated hysteresis loops with increasingly higher fields, however, increases on decreasing the temperature, but with a distinct change of slope around 165K which is close to $T^{//}_{m2}$ in Fig. 1(b). The coercive field ($E_c$) at constant electric field also starts collapsing below 165K. However, the coercive field ($E_{cS}$), obtained with a higher electric field required for saturated loops at each temperature, increases with decreasing temperature with a change of slope around 160K. The collapse of $P_r$ below 225K correlates well with the appearance of the monoclinic phase and its mesoscopic domains. The constant electric field of 12kV/cm is apparently inadequate in aligning these mesoscopic domains against the thermal fluctuations, leading to the gradual collapse of the hysteresis loop. On application of higher fields, the mesoscopic domains could also get aligned with applied electric field giving rise to saturated hysteresis loops. Hysteresis loop measurements thus clearly suggest that the low temperature monoclinic phase is a non-ergodic relaxor ferroelectric phase but with mesoscopic domains.

Jin *et al* [36] have derived the following relationship between the cell parameters of the tetragonal and low temperature monoclinic Pm phases for such a miniaturized domain state.

$$c_t = c_m - b_m + a_m \text{ and } \quad b_m = a_t, \tag{4}$$

where $c_t$ and $a_t$ are the lattice parameters of the tetragonal phase and $a_m$, $b_m$, and $c_m$ of the monoclinic Pm phase. Figure 11 shows the temperature dependence of the lattice parameters for PNN-0.35PT obtained by Rietveld analysis of XRD data collected at various temperatures while heating from 15K to 300K. It can be seen from this figure that the calculated $c_t$ (using Eq.4) in the stability field of the monoclinic phase is continuous extension of the $c_t$ measured in the stability field of tetragonal phase. Similarly, the lattice parameter $b_m$ of the monoclinic phase is a continuous extension of $a_t$. This confirms that the Jin *et al*'*s* orientation relationship [35] is fully obeyed in PNN-0.35PT for the tetragonal to monoclinic phase transition and the monoclinic domains may be visualized as miniaturization of the tetragonal domains. Both, the collapse of $P_r$ below 225K and the obeyance of the Jin *et al's* model by the cell parameters strongly support the appearance of mesoscopic domains below 225K which start freezing out around 160K giving rise to the peak in the real part of the dielectric constant around this temperature.

**4. Conclusions:**

Dielectric measurements on PNN-0.35PT in the temperature range 90 to 470K reveal a relaxor ferroelectric transition around 350K with a Vogel-Fulcher freezing temperature $T_{VF} \approx 338K$. The room temperature non-ergodic relaxor ferroelectric phase shows characteristic tetragonal splitting of the peaks in the powder XRD data. On cooling, this non-ergodic relaxor phase reenters into another ergodic relaxor phase as evidenced by the appearance of a peak in $\varepsilon^{//}(T)$ around 160K with frequency dispersion characteristic of relaxors. Temperature dependent magnetization measurements reveal that this low temperature dielectric anomaly is not linked with any magnetic transition. Low temperature XRD and elastic modulus studies reveal a structural phase transition from room temperature tetragonal to a low temperature monoclinic phase in the Pm space group around 225K. The appearance of the monoclinic phase and its new domain structure follows Jin *et al's* orientation relationship of conformally miniaturized domains of the parent tetragonal phase. As a result of the appearance of new mesoscopic domains, the constant field hysteresis loop starts collapsing below the tetragonal to monoclinic transition with a concomitant decrease in the remanent polarization. We believe that the frequency dependent dielectric relaxation peak around 160K is related with the freezing of the mesoscopic monoclinic domains, which behave as superparaelectric clusters.

**Acknowledgement:**

Satendra Pal Singh acknowledges support from the All India Council of Technical Education (AICTE) in carrying out this research.

**References:**

[1] W. Prellier, M.P.Singh and P. Murugavel, J. Phys. Condens. Matter **17,** R 803 (2005); N. Hur, S.Park, P.A. Sharma, J.S. Ahn, S. Guha and S-W. Cheong, Nature **429**, 392 (2004).

[2] G.A. Smolenskii, A.I. Agranovskaya, Sov. Phys.-Tech. Phys., 1380 (1958); H.J. Fan, M.H. Kuok, S.C. Ng, A. Yasuda, H. Ohwa, M. Iwata, H. Orihara and Y. Ishibashi, J. Appl. Phys. **91**, 2262 (2002).

[3] T. Shirakami, M. Mituskawa, T. Imai, and K.Urabe, Jpn. J. Appl. Phys. **39**, L678 (2000).

[4] T.R. Shrout J.P. Jang, KN Kim and S. Markgraf, Ferroelctric Lett. **12**, 63 (1990).

[5] J. Kuwatta, K.Uchino and S. Nomura, Ferroelctrics **37**, 579 (1981).

[6] B. Jaffe, W.R. Cook and H. Jaffe, *Piezoelectric ceramics* (Academic press London), (1971).

[7] Z. Yu, C. Ang, E. Furman and L.E. Cross, Appl. Phys. Lett. **82,** 790 (2003).

[8] P. Bao, F.Yan, Y. Dai, J. Zhu, Y. Wang and H. Luo, Appl. Phys. Lett. **84,** 5317 (2004).

[9] S.Priya, D. Viehland and K. Uchino, Appl. Phys. Lett. **80,** 4217 (2002).

[10] M. H. Lente, A. L. Zanin, E. R. M. Andreeta, I. A. Santos, D. Garcia and J. A. Eiras, Appl. Phys. Lett. **85**, 982 (2004).

[11] R. Guo, A. S. Bhalla, C. A. Randall and L.E. Cross, J. Appl. Phys. **67**, 6405 (1990).

[12] D. Viehland , J. Appl. Phys. **88**, 4794 (2000).

[13] Z. Li, L. Zhang, X. Yao, J. Mater. Res. **16**, 834 (2001).

[14] C. Lei, K. Chen, X. Zhang and J. Wang , Solid State commun. **123**, 445 (2002)

[15] L .E. Cross, Ferroelectrics, **76,** 241 (1987); L. E. Cross, Ferroelectrics, **151,** 305

(1994).

[16] D. Pandey, Key Eng. Mater. **101-102**, 177 (1995).

[17] A. Levstik, Z. Kutnjak, C.Filipič, and R. Pirc, Phys. Rev. B **57**,11204 (1998).

[18] O. Bidault, M. Licheron, E. Husson, G. Calvarin and A. Morell, Solid State commun. **98**, 765 (1996).

[19] A. K. Singh, D. Pandey and O. Zaharko, Phys. Rev.B **74,** 024101 (2006).

[20] N. Mathan, E. Husson, G. Calvarint, J.R. Gavarri, A. W. Hewat and A.Morell, J. Phys. Condens. Matter **3**,8159 (1991).

[21] Y. Yang, J. M. Liu, H. B. Huang, W. Q. Zou, P. Bao, and Z. G. Liu, Phys. Rev. B **70**, 132101 (2004).

[22] Z. J. Huang, Y. Cao, Y. Y. Sun, Y.Y. Xue, and C. W. Chu, Phys. Rev. B **56** 2623 (1997).

[23] Ragini, S.K. Mishra, D. Pandey, H. Lemmens, and G. VanTendeloo, Phys. Rev. B **64,** 054101 (2001).

[24] B. Noheda, J.A. Gonzalo, L. E. Cross, R. Guo, S.-E. Park, D. E. Cox and G., Shirane, Phys. Rev. B **61**, 8687 (2000); R. Ranjan, A.K. Singh, Ragini and D. Pandey, Phys. Rev. B**71**, 092101 (2005).

[25] B.Noheda, D.E. Cox, G. Shirane, J. Gao, and Z.-G. Ye, Phys. Rev. B **66**,054104 (2002).

[26] J. M. Kiat, Y. Uesu, B.Dkhil, M.Matsuda, C. Malibert, and G. Calvarin, Phys. Rev. B **65**, 064106 (2002).

[27] J. Rodriguez-Carvajal, Laboratory Leon Brillouin (CEA-CNRS) CEA/Saclay, 91191 Gif sur Yvette Cedex, France, Fullprof (version May 2006)

[28] D. Vanderbilt and M. H. Cohen, Phys. Rev. B **63** 094108 (2001)

[29] B. Noheda, D. E. Cox, G. Shirane, R. Guo, B. Jones and L. E. Cross, Phys.

Rev. B **63** 014103 (2000).

[30] Ragini, R. Ranjan, S. K. Mishra, and D. Pandey, J. Appl. Phys. **92** 3266 (2002).

[31] A. K. Singh and D. Pandey, Phys. Rev. B **67** 064102 (2003).

[32] S. P. Singh, A. K. Singh, and D. Pandey, J. Phys.: Condens. Matter **19**, 036217 (2007).

[33] R. J. Hill and H. D. Flack, J. Appl. Cryst. **20**, 356 (1987).

[34] R. A. Young, "The Rietveld Method" 1996 International Union of Crystallography Oxford university press, New York pp. 52-54.

[35] S. P. Singh, D. Pandey, S. Yoon, N. Shin and S.Baik (2005) (to be published).

[36] Y. M. Jin, Y. U. Wang, A.G. Khachaturyan, J. F. Li, and D. Viehland, Phys. Rev. Lett. **91**, 197601 (2003).

**TABLE I**. Refined structural parameters of the tetragonal and monoclinic phases of PNN-0.35PT at 300 and 15K, respectively. At 15K, the weight fraction of the coexisting tetragonal phase is ~ 0.40. Anisotropic thermal parameters were found necessary for $Pb^{2+}$.

| **Tetragonal (P4mm) at 300K**<br>a = b = 3.99346(8) Å; c = 4.0094(1) Å | | | | |
|---|---|---|---|---|
| Ions | x | y | z | B(Å$^2$) |
| $Pb^{+2}$ | 0.0000 | 0.0000 | 0.0000 | $\beta_{11}$ = 0.0543(7)<br>$\beta_{22}$ = 0.0543(7)<br>$\beta_{33}$ = 0.049 (1) |
| $Ti^{+4}/Ni^{+2}/Nb^{+5}$ | 0.5000 | 0.5000 | 0.525(1) | B = 0.81(5) |
| $O^{-2}_{I}$ | 0.5000 | 0.5000 | 0.043(5) | B = 0.9 (5) |
| $O^{-2}_{II}$ | 0.5000) | 0.0000 | 0.578(2) | B = 0.1(2) |
| $R_p$ = 12.3; $R_{wp}$ = 14.1; $R_{exp}$ = 9.65; $\chi^2$ = 2.13 | | | | |
| **Monoclinic (Pm) at 15K**<br>$a_m$ = 3.9946(1) Å; $b_m$ = 3.9824(2) Å; $c_m$ = 4.0129(2) Å;<br>β= 90.129(7) (degrees) | | | | |
| $Pb^{+2}$ | 0.0000 | 0.0000 | 0.0000 | $\beta_{11}$ = 0.049(3)<br>$\beta_{22}$ = 0.066(5)<br>$\beta_{33}$ = 0.058(3)<br>$\beta_{13}$ = 0.001(5) |
| $Ti^{+4}/Ni^{+2}/Nb^{+5}$ | 0.469(2) | 0.5000 | 0.538(2) | B = 0.13(1) |
| $O^{-2}_{I}$ | 0.43(1) | 0.5000 | 0.04(1) | B = 0.10(2) |
| $O^{-2}_{II}$ | 0.53(1) | 0.0000 | 0.53(2) | B = 0.18(2) |
| $O^{-2}_{III}$ | -0.02(1) | 0.5000 | 0.57(1) | B = 0.12(2) |
| $R_p$ = 10.2; $R_{wp}$ = 12.9; $R_{exp}$ = 9.17; $\chi^2$ = 1.99 | | | | |

**Figure captions:**

**Fig. 1.** Variation of the real ($\varepsilon^{/}$) and imaginary ($\varepsilon^{//}$) parts of the dielectric constant at various frequencies: [(1), 100Hz, (2) 300Hz, (3) 500Hz, (4)700Hz, (5) 1kHz, (6)3kHz, (7) 5kHz, (8) 7kHz and (9) 10kHz] in the temperature range (a) 90 K to 470K and (b) 90 K to 300K. Inset (i) in Fig. 1(a) shows the zoomed portion of the real part of the dielectric constant near the peak. Inset (ii) in Fig. 1(a) and inset to Fig. 1(b) shows the results for heating and cooling cycles at 10kHz. The arrow indicates the sense of shift of peak positions with increasing measuring frequency.

**Fig. 2.** Vogel-Fulcher fit for relaxation time corresponding to (a) the high temperature dielectric anomaly and (b) the low temperature dielectric anomaly. The nonlinear nature of the $\ln\tau$ vs. $1/T$ plot in the inset clearly rule out Arrhenius behaviour.

**Fig. 3** Variation of magnetic the susceptibility ($\chi$) with temperature for PNN-0.35PT and PNN. Inset shows the variation of $1/\chi$ with temperature, which gives negative Curie-Weiss temperatures for both PNN-0.35PT and PNN.

**Fig. 4** Variation of elastic modulus (E) with temperature for PNN-0.35PT ceramic**.** Inset shows the temperature dependence of $1/\varepsilon^{/}(T)[d(\varepsilon^{/}(T))/dT]$ with an anomaly around 225K.

**Fig. 5.** Evolution of the powder x-ray diffraction profiles of the 200, 220 and 222 pseudocubic reflections with temperature for PNN-0.35PT after stripping $CuK\alpha_2$ contribution.

**Fig. 6.** Observed (dots), calculated (continuous line), and difference (bottom line) powder diffraction profiles of the 200, 220 and 222 pseudocubic reflections obtained from full pattern Rietveld analysis of PNN-0.35PT in the 2θ range 20 to 120 degrees using different structural models : (a) tetragonal P4mm at 300K, (b) same at 15K, (c) monoclinic Cm and tetragonal coexistence, and (d) monoclinic Pm and tetragonal coexistence. The tick marks above the difference plot show the positions of the Bragg peaks. In Fig. (c) and (d) upper and lower tick marks are for tetragonal and monoclinic phases respectively.

**Fig.7.** The observed (dots), calculated (continuous line), and difference (bottom line) profiles obtained from the Rietveld refinement of PNN-0.35PT using coexistence of monoclinic phase in the Pm space group with the tetragonal phase at 15K. Inset shows the fit for higher 2θ range. The upper and lower vertical tick marks denotes the positions of the Bragg peaks for the tetragonal and monoclinic phases, respectively.

**Fig.8.** Temperature dependence of the P-E hysteresis loop with temperature at constant electric field 12kV/cm. (Scaling y axis: 1 big div. = 9μC/cm$^2$)

**Fig.9.** Saturated P-E hysteresis loop at 123K at an electric field of 22kV/cm.

**Fig.10.** The temperature variation of the remanent polarization and coercive field: The $P_r$ and $E_c$ correspond to constant field (12kV/cm) measurements whereas $P_{rS}$ and $E_{cS}$ are for higher fields required to get saturated hysteresis loops below 225K.

**Fig.11.** Variation of the lattice parameters $a_t$, $c_t$ and $a_m$, $b_m$, $c_m$, of the tetragonal and monoclinic phases with temperature for PNN-0.35PT obtained by Rietveld refinement. The lattice parameter values of $c_t$ shown with stars below 225K in the stability field of the monoclinic Pm phase correspond to those obtained using Jin *et al*'s model [36]. The vertical line marks the tetragonal to monoclinic phase transition temperature.

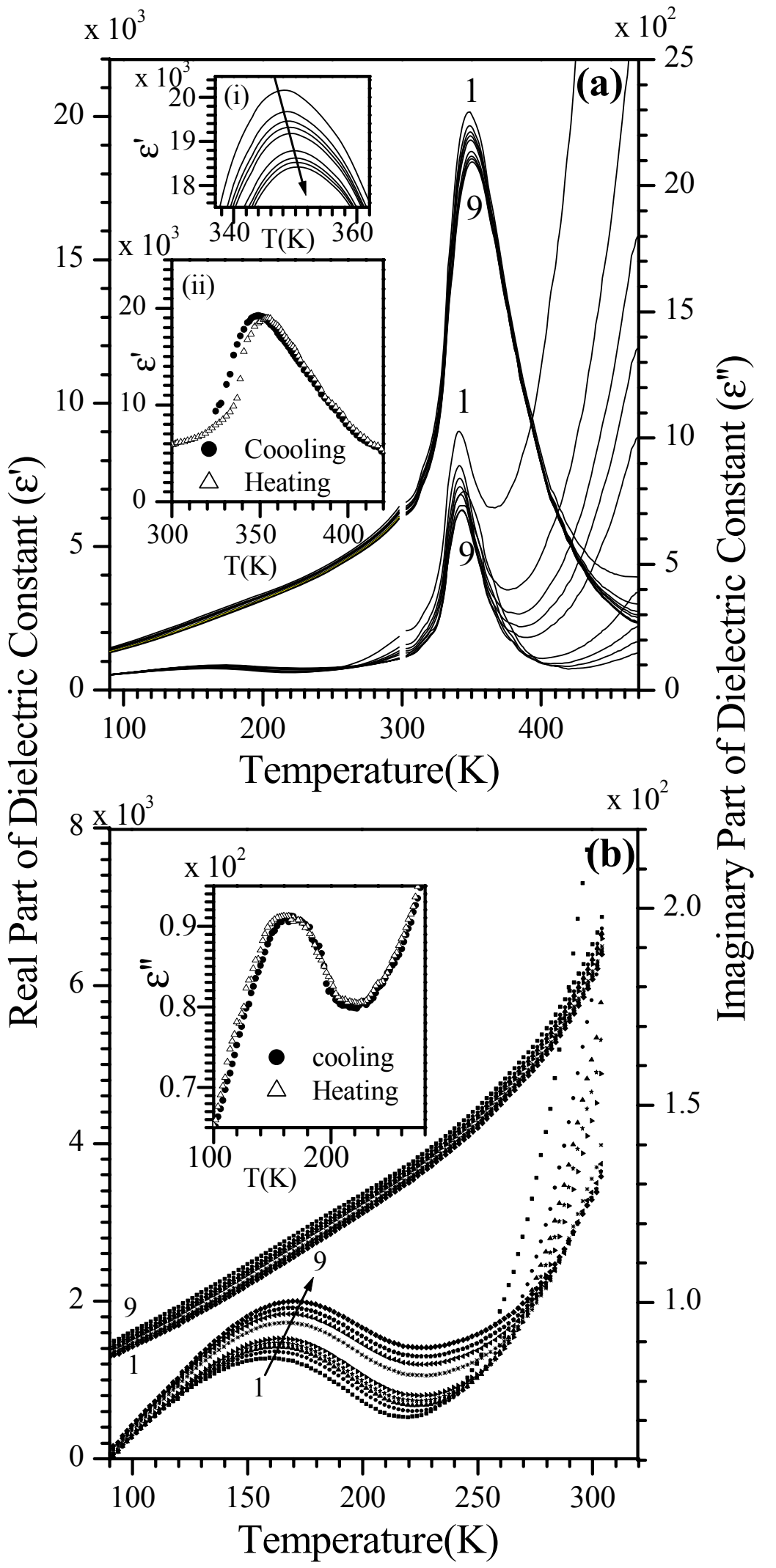

**Fig. 1**

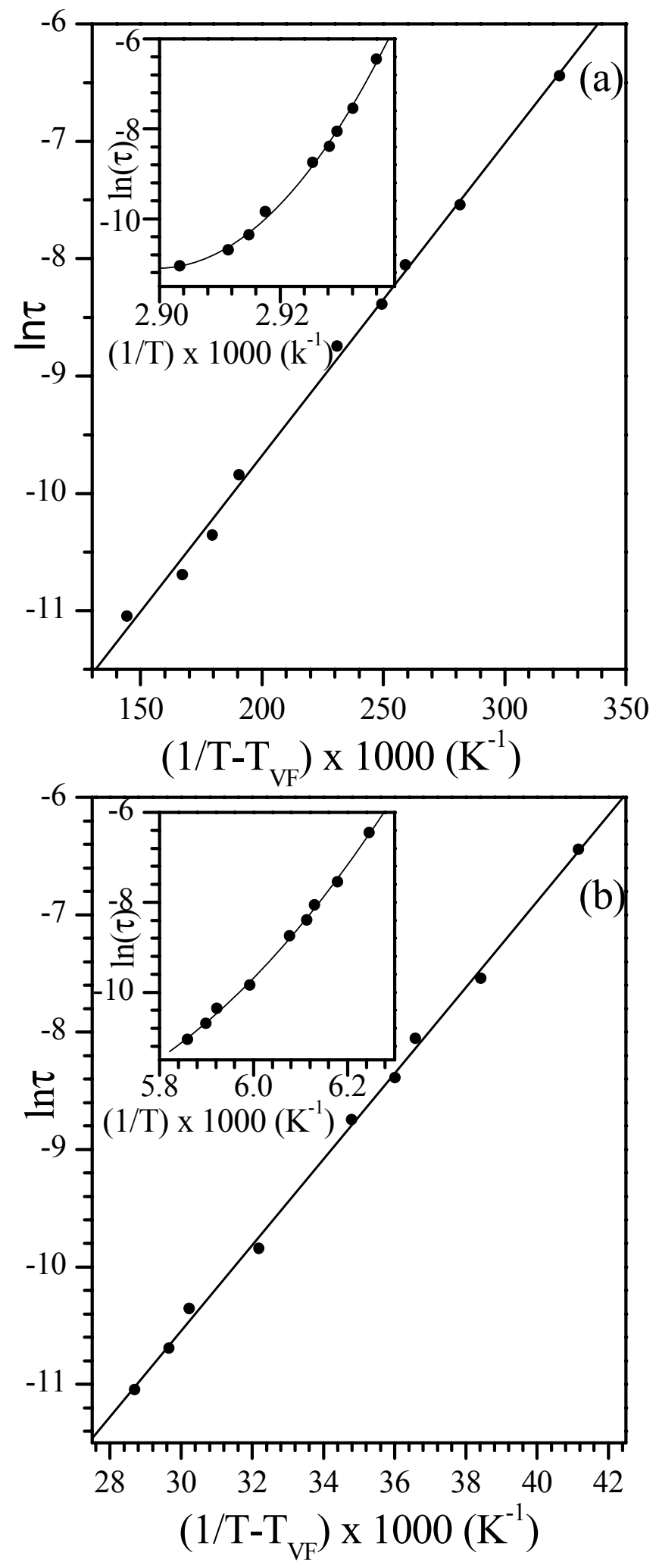
(a)
lnτ
(1/T-T$_{VF}$) x 1000 (K$^{-1}$)
ln(τ)
(1/T) x 1000 (k$^{-1}$)
(b)
lnτ
(1/T-T$_{VF}$) x 1000 (K$^{-1}$)
ln(τ)
(1/T) x 1000 (K$^{-1}$)

**Fig. 2**

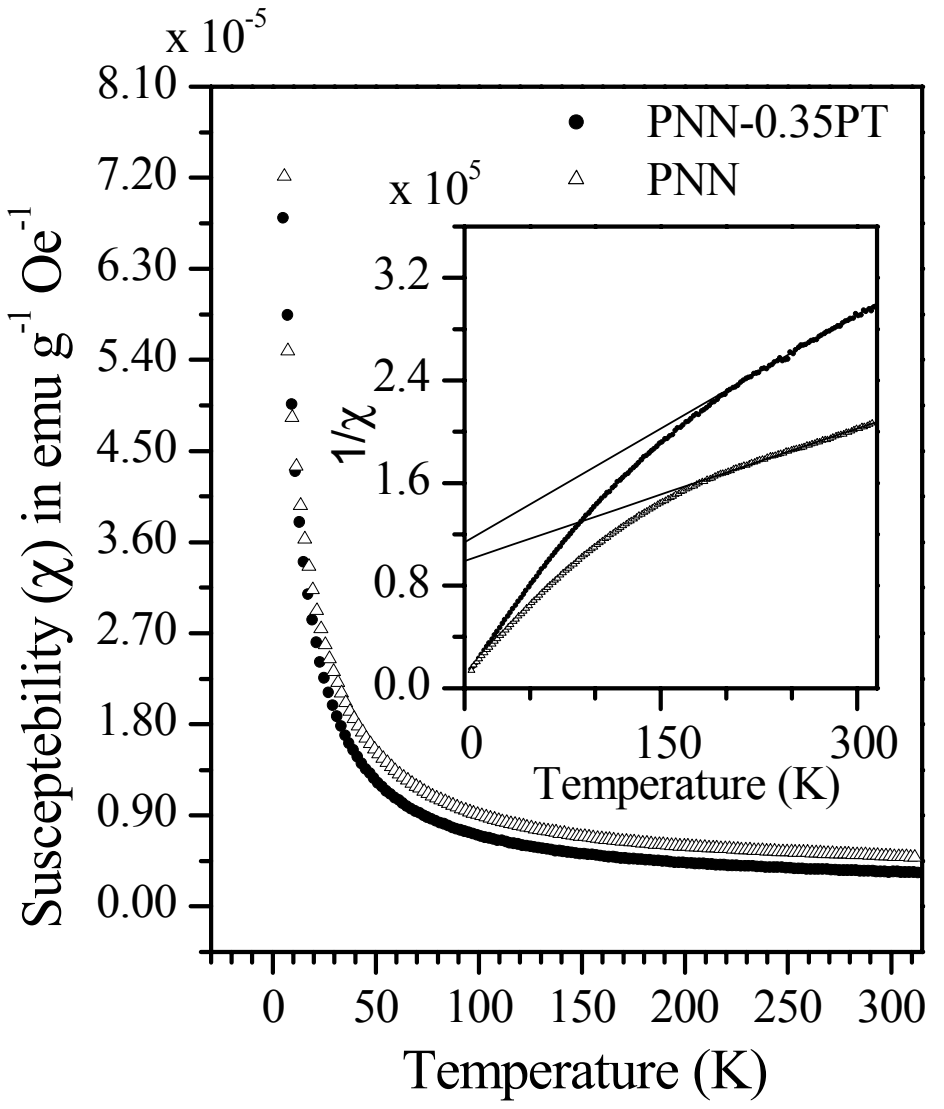

**Fig. 3**

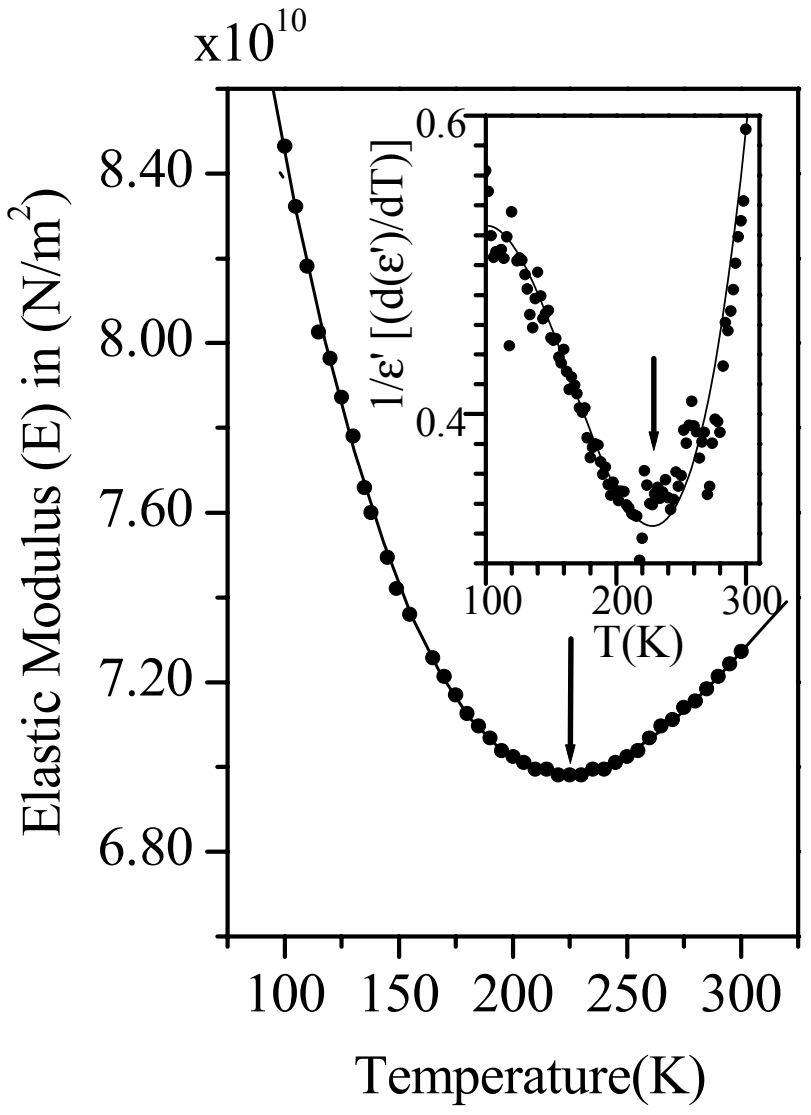

**Fig.4**

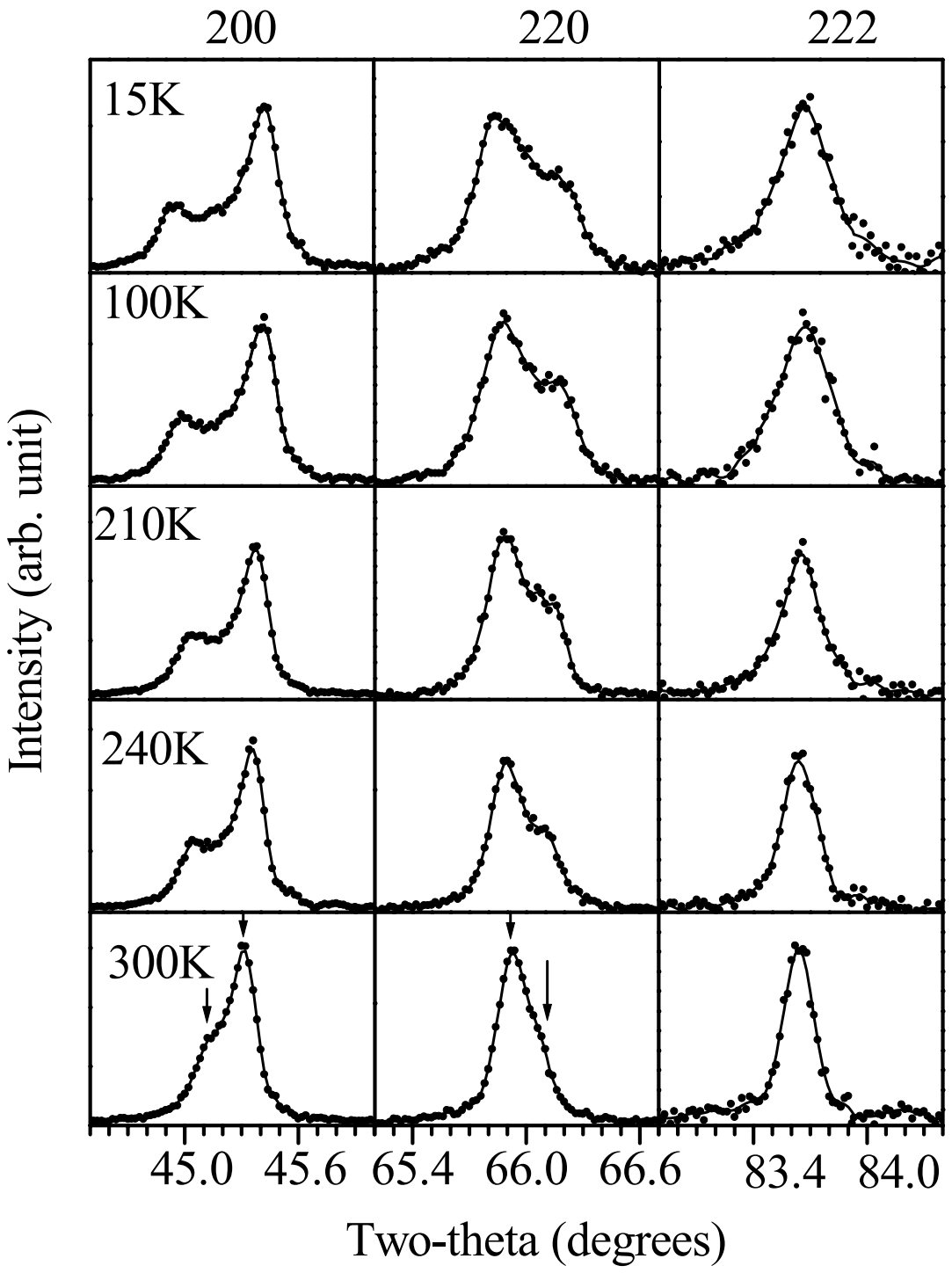

**Fig. 5**

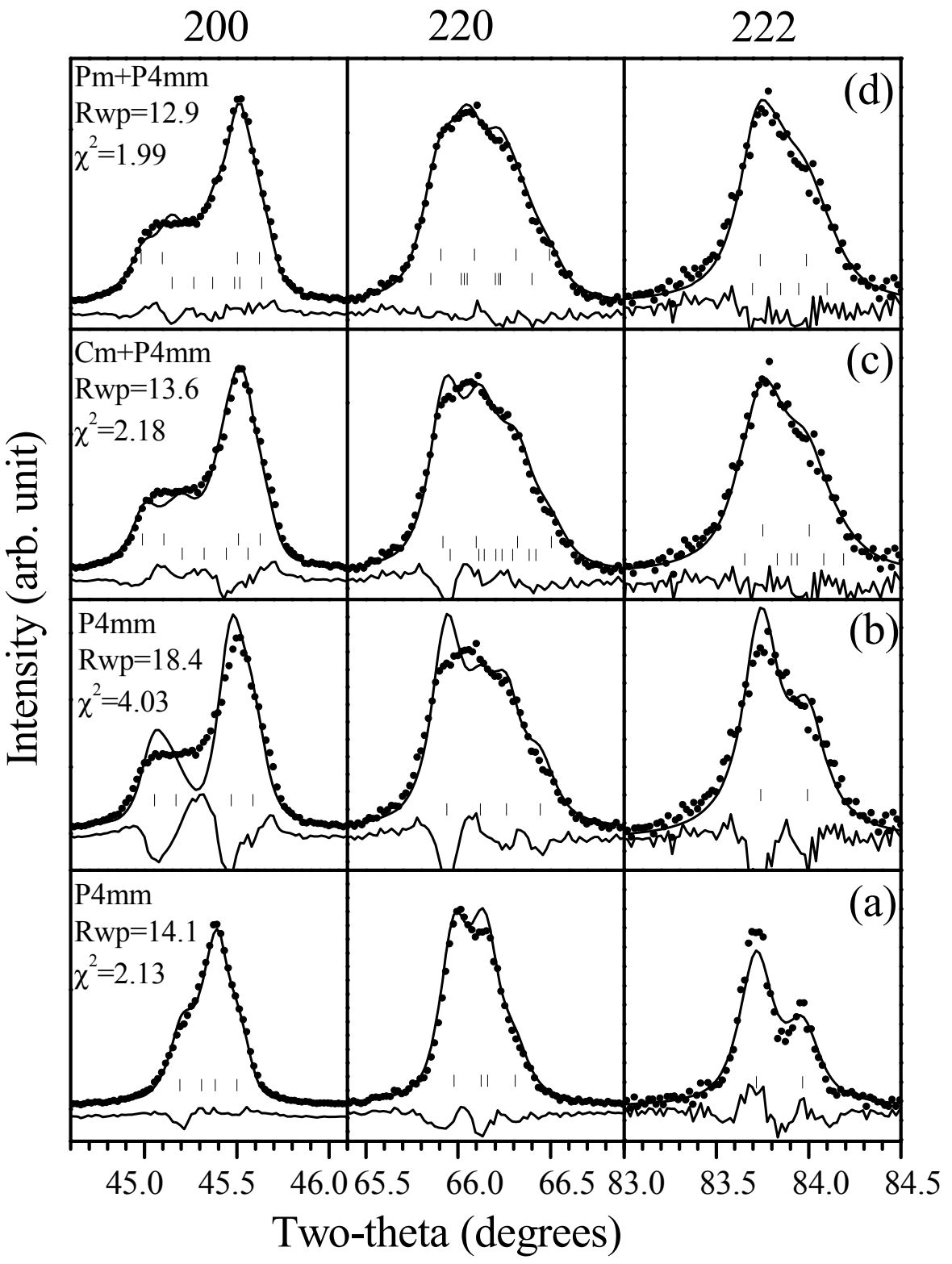
200
220
222
Pm+P4mm
Rwp=12.9
χ2=1.99
(d)
Cm+P4mm
Rwp=13.6
χ2=2.18
(c)
P4mm
Rwp=18.4
χ2=4.03
(b)
P4mm
Rwp=14.1
χ2=2.13
(a)
Intensity (arb. unit)
45.0 45.5 46.0 65.5 66.0 66.5 83.0 83.5 84.0 84.5
Two-theta (degrees)

**Fig.6**

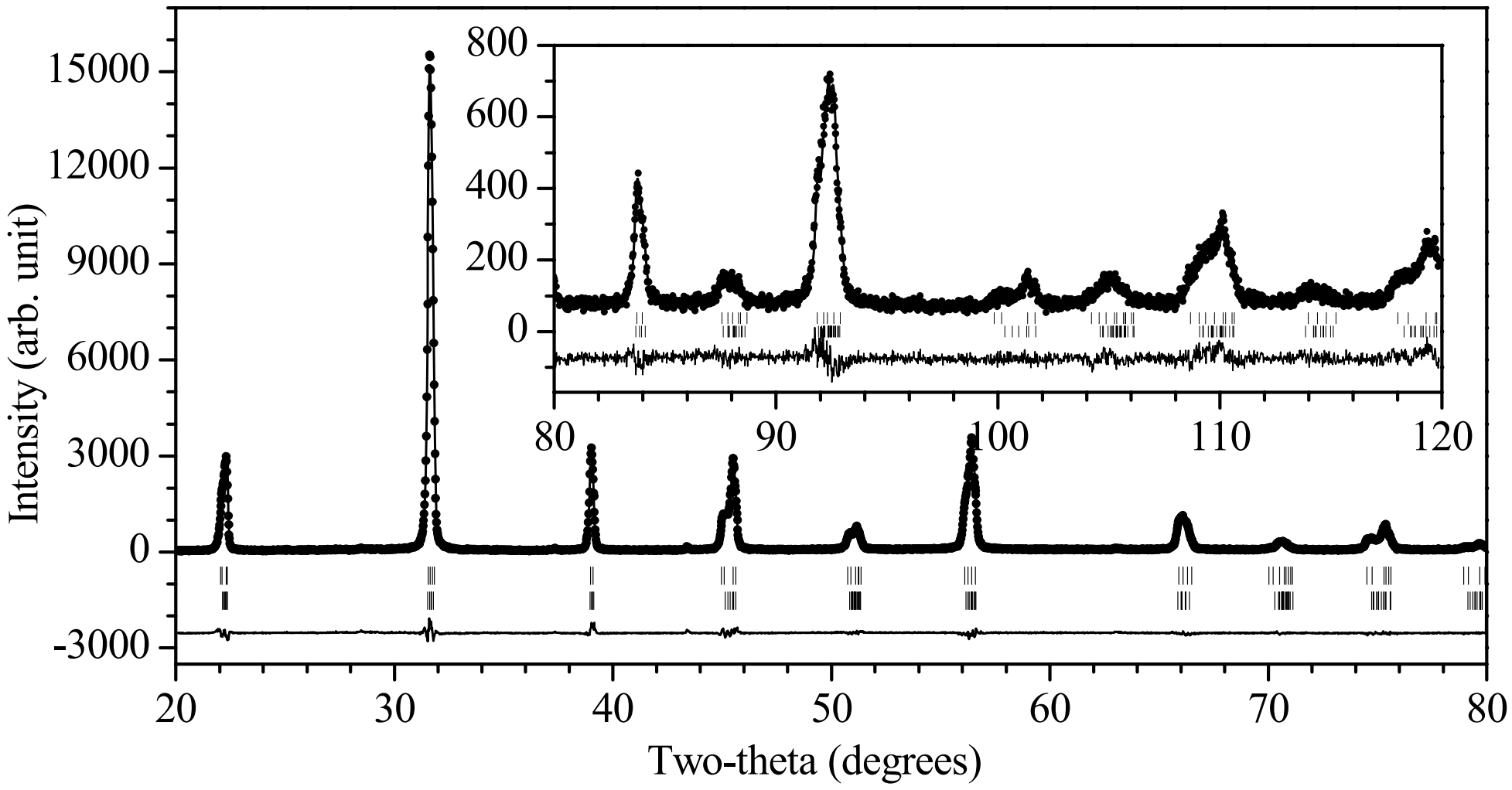

**Fig. 7**

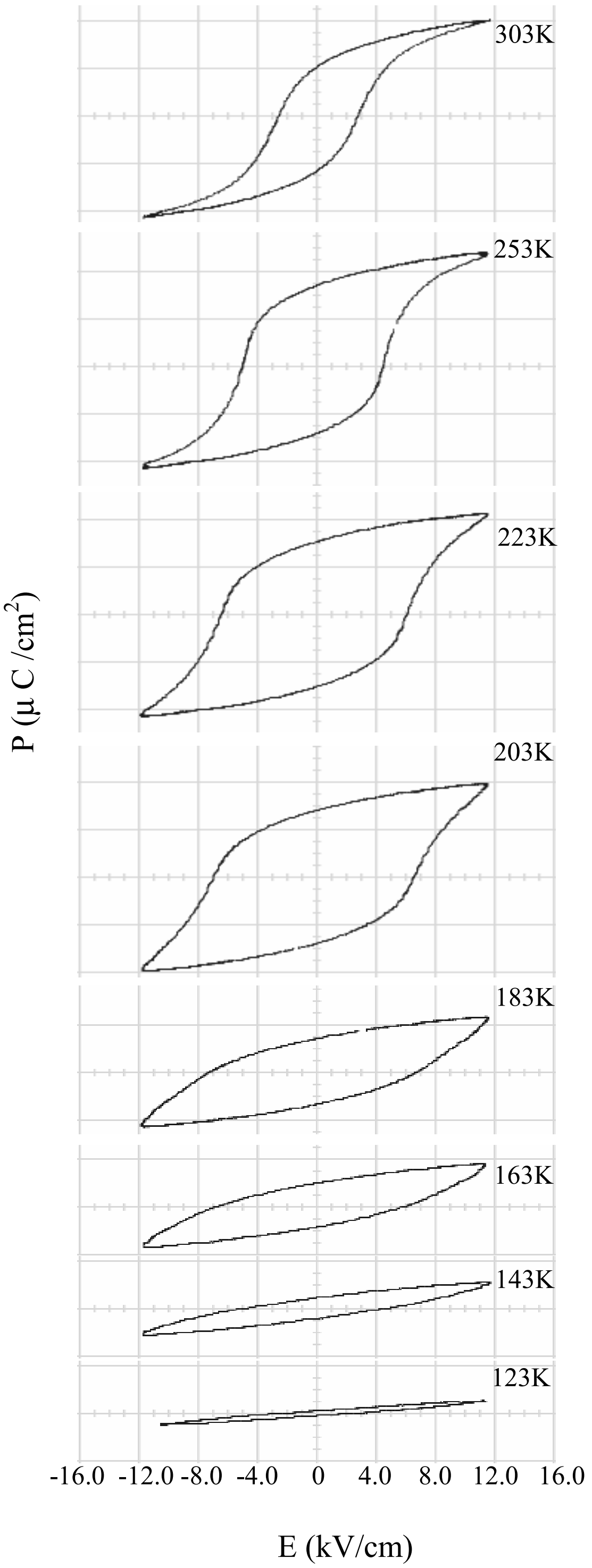
303K
253K
223K
203K
183K
163K
143K
123K
P (μ C /cm$^2$)
-16.0 -12.0 -8.0 -4.0 0 4.0 8.0 12.0 16.0
E (kV/cm)

**Fig. 8**

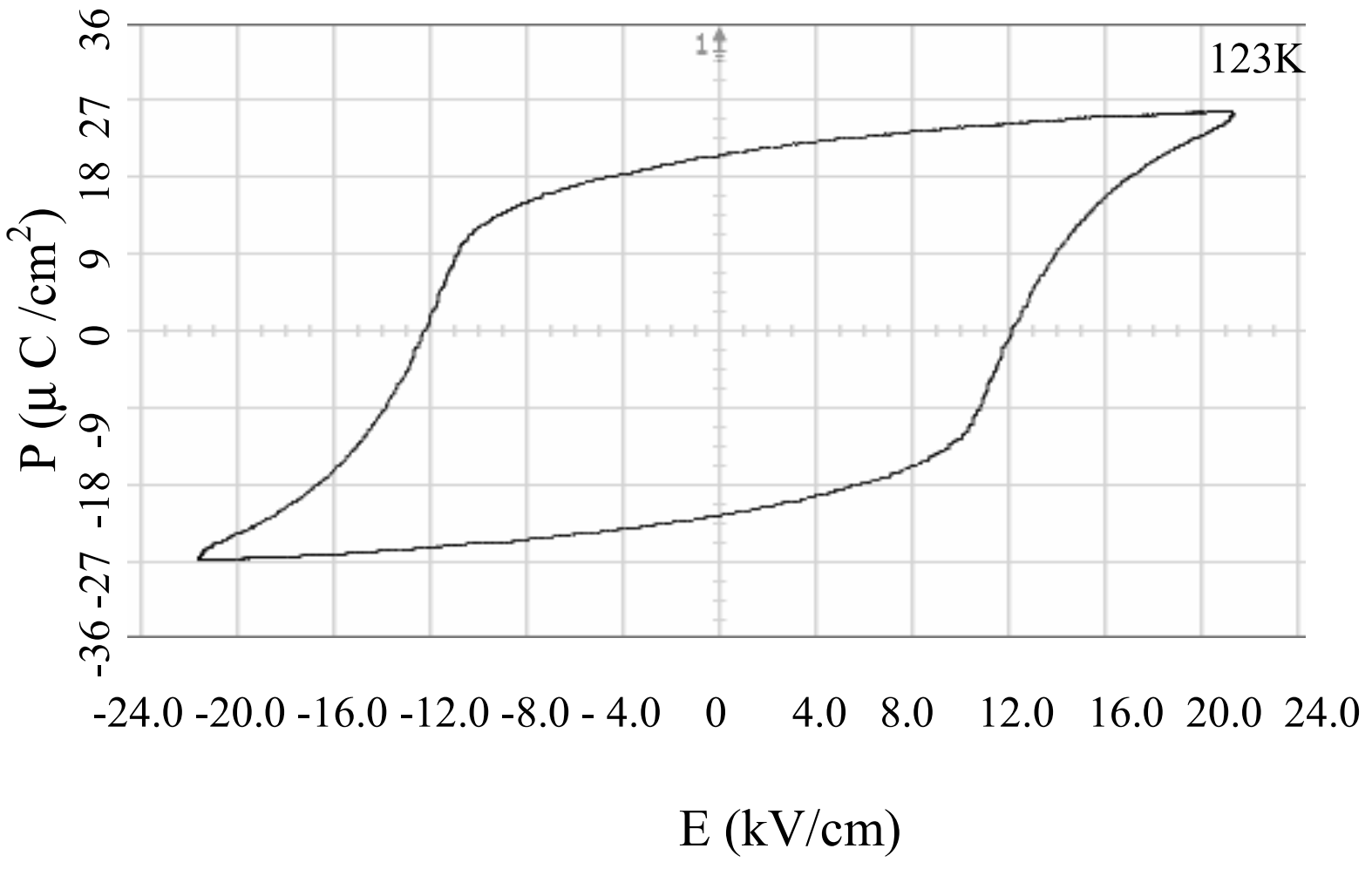
123K
P (μ C /cm2)
36
27
18
9
0
-9
-18
-27
-36
-24.0 -20.0 -16.0 -12.0 -8.0 - 4.0 0 4.0 8.0 12.0 16.0 20.0 24.0
E (kV/cm)

**Fig. 9**

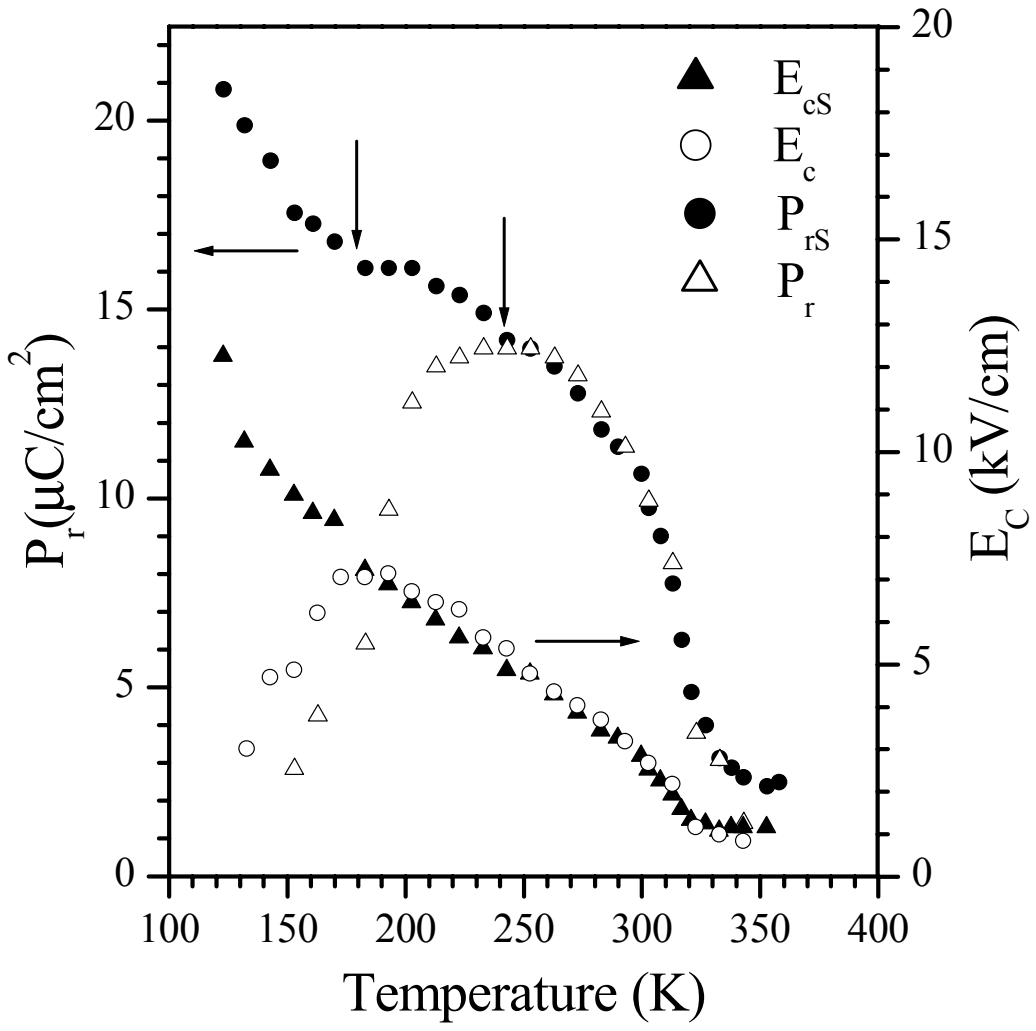

**Fig. 10**

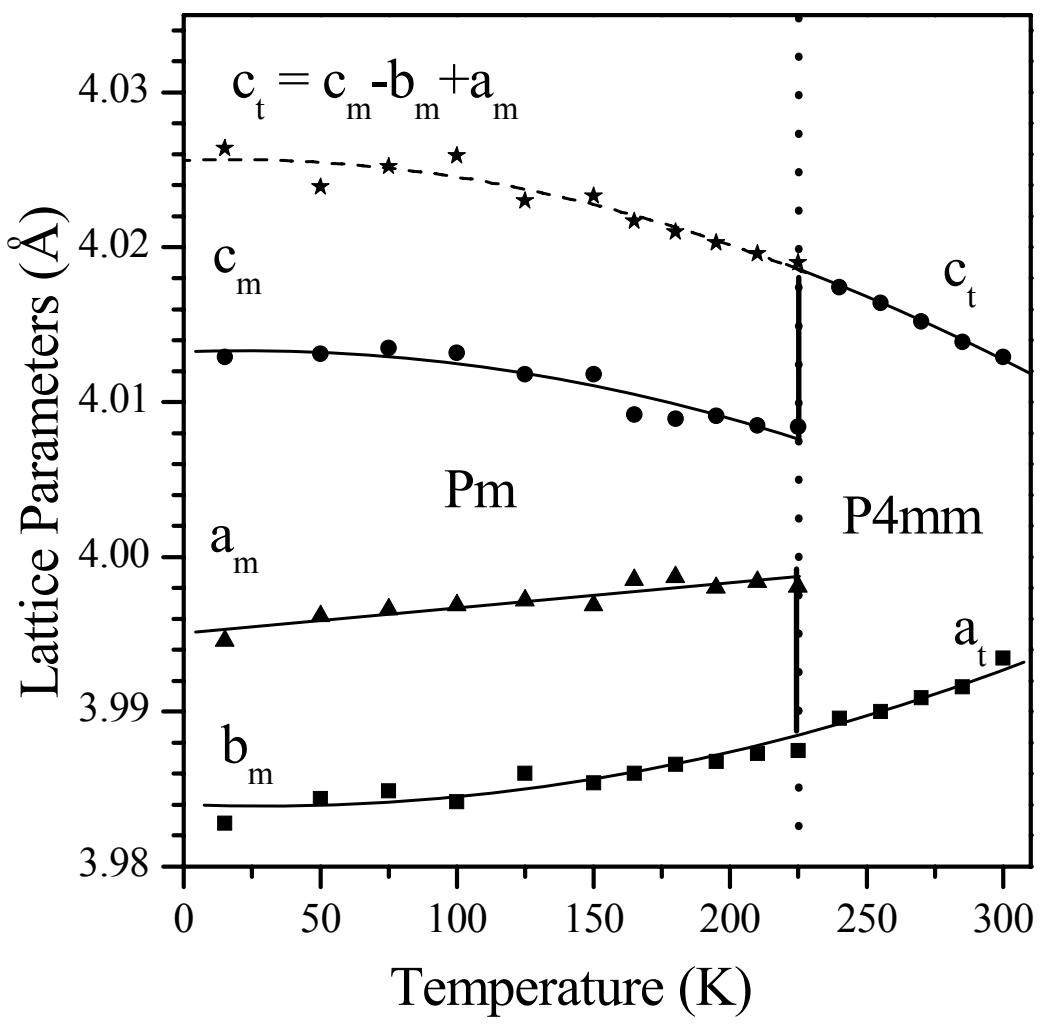

**Fig. 11**